\documentclass[final]{IEEEtran}

\usepackage{abbrevs}
\usepackage{comment}
\usepackage{flushend}
\usepackage{graphicx}
\usepackage{listings}
\usepackage{csvsimple}
\usepackage{tcolorbox}
\usepackage{todonotes}
\usepackage{tikz}
\usepackage{multirow}
\usepackage{booktabs}
\usepackage[outdir=./]{epstopdf}
\usepackage[font={footnotesize}]{caption}
\usepackage{pdfcomment}
\usepackage{relsize}
\usepackage{enumitem}
\usepackage{siunitx}
\usepackage[subtle]{savetrees}
\usepackage{microtype}
\usepackage{xspace}
\usepackage{afterpage}

\graphicspath{{figures/}}

\def\Code#1{\texttt{#1}}

\newcommand{\ropfuscator}{ROPFuscator}

\title{ROPfuscator: Robust Obfuscation with ROP}

\author{
    \IEEEauthorblockN{
        Giulio De Pasquale\IEEEauthorrefmark{1}\footnote{Equal contribution}\IEEEauthorrefmark{2}\thanks{* Equal contribution}, 
        Fukutomo Nakanishi\IEEEauthorrefmark{1}\IEEEauthorrefmark{3}, 
        Daniele Ferla\IEEEauthorrefmark{4}, 
        Lorenzo Cavallaro\IEEEauthorrefmark{5}
    }
    
    \IEEEauthorblockA{
        \IEEEauthorrefmark{2}King's College London
    }
    \IEEEauthorblockA{
        \IEEEauthorrefmark{3}Toshiba Corporation
    }
    \IEEEauthorblockA{
        \IEEEauthorrefmark{4}Università di Bologna
    }
    \IEEEauthorblockA{
        \IEEEauthorrefmark{5}University College London
    }
}

\begin{document}

\maketitle
\IEEEpeerreviewmaketitle

\begin{abstract}
  Software obfuscation is crucial in protecting intellectual property in software from reverse engineering attempts. While some obfuscation techniques originate from the obfuscation-reverse engineering arms race, others stem from different research areas, such as binary software exploitation.

  Return-oriented programming (ROP) became one of the most effective exploitation techniques for memory error vulnerabilities. ROP interferes with our natural perception of a process control flow, inspiring us to repurpose ROP as a robust and effective form of software obfuscation.
  Although previous work already explores ROP's effectiveness as an obfuscation technique, evolving reverse engineering research raises the need for principled reasoning to understand the strengths and limitations of ROP-based mechanisms against man-at-the-end (MATE) attacks.

  To this end, we present \ropfuscator{}, a compiler-driven obfuscation pass based on ROP for any programming language supported by LLVM. We incorporate opaque predicates and constants and a novel instruction hiding technique to withstand sophisticated MATE attacks. More importantly, we introduce a realistic and unified threat model to thoroughly evaluate \ropfuscator{} and provide principled reasoning on ROP-based obfuscation techniques that answer to code coverage, incurred overhead, correctness, robustness, and practicality challenges. The project's source code is published online to aid further research.
\end{abstract}

\IEEEpeerreviewmaketitle

\section{Introduction}
\label{sec:introduction}

Software has been transforming the fabric of our society for decades. It is now virtually impossible to imagine any activity that does not involve software components. As such, software is widely recognized as an important intellectual property to protect from reverse engineering attempts~\cite{aucsmith_tamper_1996,chow_approach_2001,linn_obfuscation_2003}.

In this context, obfuscation represents the de-facto standard for protecting software from being disclosed, making reverse engineering prohibitively expensive, i.e., allowing a vendor to collect enough
revenue at the peak of a marketing campaign. However, although useful, there is no clear winner in the obfuscation-reverse engineering arms race~\cite{schrittwieser_protecting_2016}. Thus, relying on an arsenal of obfuscation techniques seems to be the only effective way to counteract attempts to break such confidentiality requirements. Dummy code insertion \cite{linn_obfuscation_2003}, control flow flattening \cite{chow_approach_2001}, self-modifying code \cite{kanzaki_exploiting_2003}, opaque predicates \cite{collberg_manufacturing_1998}, virtualization \cite{anckaert_proteus_2006}, anti-debugging~\cite{gagnon_software_2007} and mixed-boolean arithmetic (MBA)~\cite{schloegel2022loki,eyrolles2017obfuscation,zhou2007information} are well-known software obfuscation techniques. They all exploit assumptions that challenge---one way or another---the logic of reverse engineering algorithms. For instance, inserting dummy code and opaque predicates interferes with attempting to reconstruct an underlying algorithm's semantics directly. In contrast, control flow flattening breaks the ability to understand a program's execution flow, challenging further reasoning to identify properties of interest.

While advances in reverse engineering spin the creation of more sophisticated obfuscation techniques, others result from intriguing leaps from different research areas. In this context, return-oriented programming (ROP) gained popularity as one of the most advanced memory error exploitation techniques~\cite {shacham_geometry_2007}. Core to this is the ability to chain the invocation of chunks of code (gadgets) to execute arbitrary - and often malicious - code. As such, ROP builds its working logic on threaded code~\cite{bell_threaded_1973}, changing the usual interpretation of code execution centered on the instruction pointer, for one, pivoted on the stack pointer.

This observation naturally suggests ROP can be repurposed to represent a robust and effective form of software obfuscation. First, threaded code changes our understanding of control flow graphs, de-facto breaking subsequent data-flow analysis that relies on them. Second, ROP provides fine control of the obfuscation's granularity, operating at the level of individual assembly instructions. Third, an obfuscated piece of code would see its semantics due to the execution of code gadgets scattered throughout the entire process address space.

Prior work already explores the effectiveness of ROP as an obfuscation technique. For instance, Mohan et al.~\cite{mohan_frankenstein_2012}, and Borrello et al.~\cite{borrello_rop_2019} repurpose ROP to challenge malware detection tasks. Teuwen et al.~\cite{rop_patent_2014} introduced the first patent on ROP as an obfuscation technique; Mu et al.~\cite{mu_ropob_2018} apply ROP to obfuscate a program's control flow graph at a coarse granularity, ignoring fine-grained obfuscation of individual assembly instructions. Despite being promising, they all fail to explore the assumptions and the extents to which ROP represents a viable solution for obfuscation techniques to withstand man-at-the-end (MATE) reverse engineering attacks~\cite{akhunzada_man-at--end_2015}, where reverse engineering attempts tailored at ROP-based obfuscation~\cite{lu_derop_2011,graziano_ropmemu_2016,yadegari_generic_2015} or dynamic symbolic execution~\cite{shoshitaishvili_sok_2016} still undermine its effectiveness. In another work, Borrello et al.~\cite{borrello_hiding_particles_2021} proposed a binary-rewriting approach that uses ROP chains coupled with a custom algorithm to protect its gadget addresses.

We present \ropfuscator{}, a compiler-driven obfuscation pass for any programming language supported by LLVM. At its core, \ropfuscator{} relies on ROP microgadgets~\cite{homescu_microgadgets_2012} to obfuscate arbitrary programs at the granularity of individual assembly instructions. To withstand attacks of increasing sophistication, \ropfuscator{} relies on opaque predicates and constants~\cite{moser_limits_2007} to protect the recovery of the gadget locations in the address space. We present a thorough reproducible evaluation of \ropfuscator{} across five dimensions, which supports our principled reasoning with evidence on completeness (code coverage), incurred overhead, correctness, robustness to MATE attacks, and practicality. We release \ropfuscator{} to support further the need for principled reasoning in domains characterized by an endemic attack-defense arms race.

\noindent In summary, we make the following contributions:
\begin{itemize}[topsep=1pt]
    \item We introduce a unified threat model that ROP-based obfuscation techniques must address to assess their robustness to increasingly sophisticated MATE attacks (\S\ref{subsec:obfuscation:threatmodel}). This helps us to provide principled reasoning to identify and justify the design choices that avoid brittle arms races that are endemic to the software obfuscation domain instead of providing researchers and practitioners with a clear and contextualized understanding of strengths and limitations.
    \item We present \ropfuscator{}, a compiler-driven obfuscation pass based on ROP (\S\ref{subsec:implementation:rop}) for any programming language supported by LLVM. To withstand sophisticated MATE attacks, we equip \ropfuscator{} with opaque predicates and constants (\S\ref{subsec:implementation:opaque}), and we build a novel instruction hiding technique that intertwines ROP gadget of arbitrary length in opaque predicates to challenge analysis in distinguishing between the two and thus the semantics of the obfuscated code against code to withstand analyses~(\S\ref{subsec:implementation:hiding}).
    \item We present a thorough evaluation of \ropfuscator{} along five dimensions to support our principled reasoning and provide the opportunity to understand its effectiveness in practical contexts (\S\ref{sec:evaluation}).
\end{itemize}

\newcommand{\ThreatA}{ROP-agnostic Static Analysis}
\newcommand{\ThreatB}{Static ROP Chain Analysis}
\newcommand{\ThreatC}{Dynamic Symbolic Execution}
\newcommand{\ThreatD}{Dynamic ROP Chain Analysis}
\newcommand{\ShortThreatA}{Static Analysis}
\newcommand{\ShortThreatB}{Static ROP Analysis}
\newcommand{\ShortThreatC}{DSE}
\newcommand{\ShortThreatD}{Dynamic ROP Analysis}

\section{Threat Model}
\label{subsec:obfuscation:threatmodel}

Our threat model considers MATE attacks of increasing sophistication~\cite{akhunzada_man-at--end_2015}. In particular, we assume attackers can rely on static ROP-agnostic and ROP chain disassembly analyses, dynamic symbolic execution, and ROP-specific dynamic analyses. In doing so, we adopt metrics similar to the one used in previous work on software obfuscation~\cite{linn_obfuscation_2003,ollivier_how_2019}.

For simplicity and ease, referencing these threats to motivate the design choices and support the underlying principled reasoning and evaluation, we refer to the following threats as \textbf{Threat A-D}. However, they should not be seen as individual and disconnected threat models. On the contrary, they represent a realistic and unified threat model that explores how robust \ropfuscator{} is in facing adaptive attacks that are aware of \ropfuscator{} inner working mechanisms.

\paragraph{Threat A: \ThreatA{}} The core of static analysis of binary programs is disassembly. Linear sweep and recursive traversal are two main static disassembly algorithms that aim to recover a program's assembly instructions by analyzing a sequence of bytes linearly (linear sweep) or following the expected execution flow (recursive traversal). Decompilation is often built on a successful disassembly to convert assembly code into high-level program constructs.

\paragraph{Threat B: \ThreatB{}} It is perhaps unsurprising that ROP chains' introduction in a program naturally breaks traditional disassembly algorithms. They assume a code execution model relying on an architecture-specific instruction pointer register; on the other hand, ROP, built on threaded code, reuses the stack pointer register - or, more generally, any other general-purpose register - to keep track of the next instruction to execute. A more realistic threat model here should thus consider the ability of statically analyzing ROP chains to reconstruct the original control flow of the obfuscated program, which can be leveraged to build more insightful data flow and decompilation analyses.

\paragraph{Threat C: \ThreatC{}} Static ROP chain analysis requires identifying the address of ROP gadgets. Address-agnostic ROP gadgets, therefore, challenge this analysis effectively. The mechanism to build ROP chains to hide ROP gadgets is not straightforward but is discussed thoroughly in the following sections. Here, it is enough to assume this is a possibility. Therefore, our threat model must include attacks that rely on dynamic symbolic execution (DSE) to identify such information. Once successful, one can rely on the above analyses to recover the original program's semantics.

\paragraph{Threat D: \ThreatD{}} This analysis takes advantage of runtime information in a context where the attacker knows ROP is a core building block for program obfuscation. Instruction traces collected from a running process are passed to a CPU emulator, which executes the ROP chains, extracting the original code from the gadgets~\cite{graziano_ropmemu_2016,delia_static_2019}.

\section{Architecture and Implementation}
\label{sec:implementation}

In the following sections, we present a more detailed view of \ropfuscator{}, show its obfuscation steps, and how they are interconnected. 

\begin{figure*}[t]
	\def\svgwidth{17cm}
	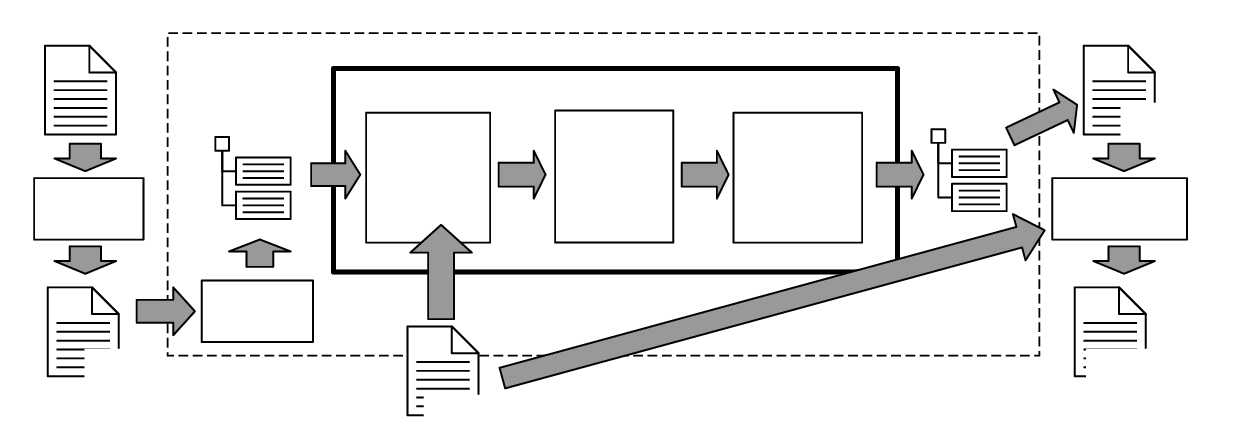
\caption{An architectural view of \ropfuscator{}.}
\label{fig:overview}
\end{figure*}

\subsection{Architectural Overview}
\label{subsec:obfuscation:architecture}

Our framework obfuscates LLVM-produced code at the x86 assembly level. 
The source code is compiled into LLVM's IR and processed by ROPfuscator. It consists of three components called in subsequent order, shown in Figure~\ref{fig:overview}, named ROP transformation, opaque predicate insertion, and instruction hiding.

The first, \textit{ROP transformation} (\S\ref{subsec:implementation:rop}) converts instructions into ROP chains, while the second one \textit{Opaque predicate insertion} (\S\ref{subsec:implementation:opaque}) injects opaque predicates in the ROP chain generation code to protect the gadgets' entry point address and other program immediates. Finally, \textit{instruction hiding} (\S\ref{subsec:implementation:hiding}) picks some instructions and embeds them into opaque predicates.

The obfuscation components can be applied selectively while respecting their invocation order. For example, ROP transformation can be applied independently, while the opaque predicates pass should be generally used after executing the ROP transformation.

\subsection{ROP transformation}
\label{subsec:implementation:rop}

Methods of converting normal code to equivalent gadgets are proposed in several studies~\cite{schwartz_q_2011,mohan_frankenstein_2012}. However, instead of processing native machine instructions, they transform various intermediate representations into ROP gadgets. 
In our work, we take a significantly different approach. As shown in Figure~\ref{fig:obfuscation-example}, \ropfuscator{} decomposes a native x86 instruction into multiple instructions, which are then matched with the available microgadgets. 
The address of the gadgets, along with the symbol table of the target library, is used to emit the obfuscated ROP code.    
We explain this process in detail in the following paragraphs.

\paragraph{Gadgets extraction}
The extraction process is based on the Galileo algorithm~\cite{shacham_geometry_2007}. The algorithm identifies \Code{ret} instructions and scans in reverse to locate a valid instruction sequence, the ROP gadget. The gadgets are extracted from a shared library chosen by the user. For design simplicity, we only rely on \emph{microgadgets}~\cite{homescu_microgadgets_2012} of length 1 (i.e, only one instruction before the \Code{ret} instruction) to build ROP chains.

\paragraph{ROP chain generation}
The use of microgadgets may incur the unavailability of gadgets needed to perform operations on registers. For this reason, we decompose the original instruction in smaller computations that use temporary registers (Step \ref{obf-step:decompose} in Figure~\ref{fig:obfuscation-example}). The temporary registers are found by performing live register analysis~\cite{tamches_fine-grained_1999} for each instruction within the basic blocks. Once the available registers are enumerated, we use gadgets to exchange them accordingly, similarly to the method proposed by Homescu et al.~\cite{homescu_microgadgets_2012}, to generate ROP chains (Step \ref{obf-step:ropchain} in Figure~\ref{fig:obfuscation-example}).

\paragraph{Emitting ROP generator code}
Once the gadgets are extracted, the ROP chain must be built and injected into the program. This is done by adding \emph{rop generator code} which pushes the generated ROP chain onto the stack in reverse order, followed by a \Code{ret} instruction (Step \ref{obf-step:ropchain-generator} in Figure~\ref{fig:obfuscation-example}).
A gadget address is calculated using the address of a random symbol from the linked library as a base address to deal with ASLR. Later, the offset of the gadget address is added to the base address and then pushed to the stack.
We do not use symbols defined in other linked libraries or the program to avoid symbol conflicts while computing the gadget addresses.

\afterpage{
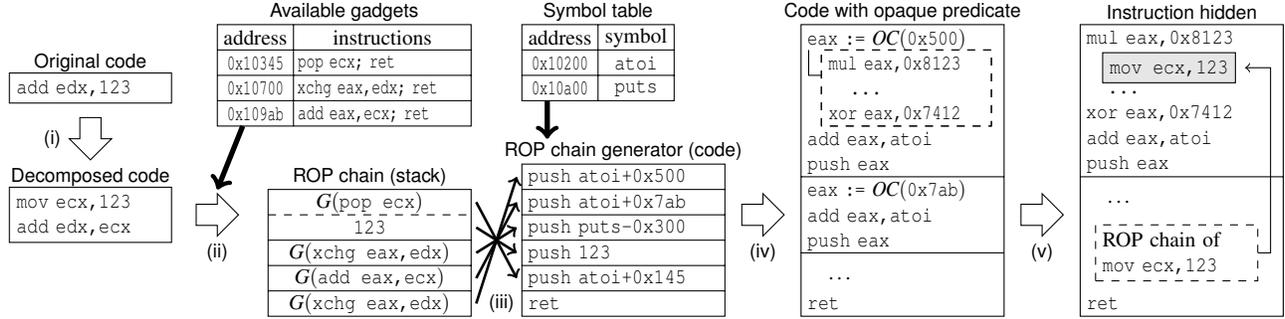
\begin{figure*}
	\centering
	\footnotesize
    \resizebox{\textwidth}{!}{
	\newcounter{stepnumvalue}
	\def\thestepnumvalue{\textsf{(\roman{stepnumvalue})}}
	\def\stepnum{\refstepcounter{stepnumvalue}\thestepnumvalue}
	\newsavebox{\boxrightarrow}
	%
	\sbox{\boxrightarrow}{
		\begin{tikzpicture}[x=0.1em,y=0.1em]
			\draw (0,5)--(0,15)--(10,15)--(10,20)--(20,10)--(10,0)--(10,5)--cycle;
		\end{tikzpicture}
	}
	\begin{tikzpicture}[x=0.12em,y=0.12em]
		\node at (30,95){\textsf{Original code}};
		\node[draw,text width=7em] at (30, 85) {\texttt{add edx,123}};
		\node at (30,65){\rotatebox{-90}{\usebox{\boxrightarrow}}};
		\node at (15,65){\stepnum\label{obf-step:decompose}};
		\node at (30,50){\textsf{Decomposed code}};
		\node[draw,text width=7em] at (30, 35) {\texttt{mov ecx,123 \\ add edx,ecx}};
		\node at (80,35){\usebox{\boxrightarrow}};
		\node at (80,20){\stepnum\label{obf-step:ropchain}};
		\draw[->,line width=2pt] (90,70)--(80,45);
		\node at (130,115){\textsf{Available gadgets}};
		\draw (80,70)--(180,70)--(180,110)--(80,110)--cycle
		(110,70)--(110,110)
		(80,100)--(180,100)
		(80,90)--(180,90)
		(80,80)--(180,80);
		\node at (95,105){address};
		\node at (145,105){instructions};
		\node at (95,95){\scalebox{.75}[1.0]{\texttt{0x10345}}};
		\node[text width=8em] at (145,95){\scalebox{.8}[1.0]{\texttt{pop ecx; ret}}};
		\node at (95,85){\scalebox{.75}[1.0]{\texttt{0x10700}}};
		\node[text width=8em] at (145,85){\scalebox{.8}[1.0]{\texttt{xchg eax,edx; ret}}};
		\node at (95,75){\scalebox{.75}[1.0]{\texttt{0x109ab}}};
		\node[text width=8em] at (145,75){\scalebox{.8}[1.0]{\texttt{add eax,ecx; ret}}};
		\node at (230,115){\textsf{Symbol table}};
		\draw (200,80)--(260,80)--(260,110)--(200,110)--cycle
		(230,80)--(230,110)
		(200,100)--(260,100)
		(200,90)--(260,90);
		\node at (215,105){address};
		\node at (245,105){symbol};
		\node at (215,95){\scalebox{.75}[1.0]{\texttt{0x10200}}};
		\node at (245,95){\texttt{atoi}};
		\node at (215,85){\scalebox{.75}[1.0]{\texttt{0x10a00}}};
		\node at (245,85){\texttt{puts}};
		\node at (140,50){\textsf{ROP chain (stack)}};
		\draw (100,-5)--(100,45)--(180,45)--(180,-5)--cycle
		(100,25)--(180,25)
		(100,15)--(180,15)
		(100,5)--(180,5);
		\draw[dashed] (100,35)--(180,35);
		\node at (140,40){$G(\texttt{pop ecx})$};
		\node at (140,30){\texttt{123}};
		\node at (140,20){$G(\texttt{xchg eax,edx})$};
		\node at (140,10){$G(\texttt{add eax,ecx})$};
		\node at (140,0){$G(\texttt{xchg eax,edx})$};
		\draw[->,line width=1pt] (182,0)--(198,50);
		\draw[->,line width=1pt] (182,10)--(198,40);
		\draw[->,line width=1pt] (182,20)--(198,30);
		\draw[->,line width=1pt] (182,30)--(198,20);
		\draw[->,line width=1pt] (182,40)--(198,10);
		\draw[->,line width=2pt] (210,80)--(210,65);
		\node at (192,0){\stepnum\label{obf-step:ropchain-generator}};
		\node at (240,60){\textsf{ROP chain generator (code)}};
		\draw (200,-5)--(200,55)--(280,55)--(280,-5)--cycle
		(200,45)--(280,45)
		(200,35)--(280,35)
		(200,25)--(280,25)
		(200,15)--(280,15)
		(200,5)--(280,5);
		\node[text width=9em] at (240,50){\texttt{push atoi+0x500}};
		\node[text width=9em] at (240,40){\texttt{push atoi+0x7ab}};
		\node[text width=9em] at (240,30){\texttt{push puts-0x300}};
		\node[text width=9em] at (240,20){\texttt{push 123}};
		\node[text width=9em] at (240,10){\texttt{push atoi+0x145}};
		\node[text width=9em] at (240,0){\texttt{ret}};
		\node at (295,35){\usebox{\boxrightarrow}};
		\node at (295,20){\stepnum\label{obf-step:opaque}};
		\node at (350,115){\textsf{Code with opaque predicate}};
		\draw (310,-5)--(310,110)--(390,110)--(390,-5)--cycle
		(310,50)--(390,50)
		(310,20)--(390,20)
		(313,100)--(313,90)--(318,90);
		\draw[dashed] (318,70)--(318,100)--(385,100)--(385,70)--cycle;
        \node[text width=9em] at (460,80){%
			\texttt{%
				mul eax,0x8123 \\
				~\\
				\hskip1em\dots \\
				xor eax,0x7412 \\
				add eax,atoi \\
				push eax
			}};
		\node[text width=9em] at (350,35){%
			\scalebox{.95}[1.0]{\texttt{eax := $OC(\texttt{0x7ab})$}} \\
			\texttt{%
				add eax,atoi \\
				push eax
			}};
		\node[text width=9em] at (350,10){\hskip1em\dots};
		\node[text width=9em] at (350,0){\texttt{ret}};
  		\node at (405,35){\usebox{\boxrightarrow}};
		\node at (405,20){\stepnum\label{obf-step:hiding}};
		\node at (460,115){\textsf{Instruction hidden}};
		\draw (420,-5)--(420,110)--(500,110)--(500,-5)--cycle
		(420,50)--(500,50);
		\draw[dashed] (318,70)--(318,100)--(385,100)--(385,70)--cycle;
		\node[text width=9em] at (350,80){%
			\scalebox{.95}[1.0]{\texttt{eax := $OC(\texttt{0x500})$}} \\
			\texttt{%
				\hskip1em \scalebox{.9}[1.0]{mul eax,0x8123} \\
				\hskip2em \dots \\
				\hskip1em \scalebox{.9}[1.0]{xor eax,0x7412} \\
				add eax,atoi \\
				push eax
			}};
        \node[text width=7em,draw,dashed] at (458,20){%
			ROP chain of \\
			\texttt{mov ecx,123}
		};
		\draw[->] (490,20)--(495,20)--(495,93)--(485,93);
		\node[draw,preaction={fill, black!10}] at (455,93){\texttt{{mov ecx,123}}};
		\node[text width=9em] at (460,40){\hskip1em\dots};
		\node[text width=9em] at (460,0){\texttt{ret}};
	\end{tikzpicture}
 }
	\caption{Obfuscation Transformation Example with ROP, Opaque Predicates, and Instruction Hiding.}
	\label{fig:obfuscation-example}
\end{figure*}
}

\paragraph{Program semantics preservation}
It is crucial to preserve the semantics of the program while obfuscating. One of the peculiarities of the x86 ISA is that it contains many instructions that can modify the flag register. Therefore, we carefully use the \Code{pushf} and \Code{popf} instructions to save and restore the flags on necessity. Fortunately, since LLVM exposes valuable metadata to check whether register flags may be safely clobbered, we can use this information to omit flag saving when unnecessary and thus drastically improve the performance.

\subsection{Opaque Predicates and Instruction Hiding}
\label{subsec:implementation:opaque}

This section will discuss how opaque predicates are generated and later inserted into the ROP generation code. Besides, we detail how static and dynamic analysis affects opaque predicates and our approach to improving their resilience.

Several opaque predicates generation algorithms have been proposed in previous works. They are based on arithmetic operations, non-determinism~\cite{collberg_manufacturing_1998}, one-way functions~\cite{zobernig_when_2019} and computationally hard programs (e.g. pointer-aliasing~\cite{collberg_manufacturing_1998}, 3SAT~\cite{moser_limits_2007,sheridan_manufacturing_2016}).

We tested the use of integer factorization and the 3SAT problem for generating opaque predicates. 

The algorithm based on \emph{integer factorization} takes two 32-bit inputs $x,y$ and returns $0$ if $xy=C$ for a fixed 64-bit prime integer $C$, returning $1$ otherwise. The factorization of $C$ is needed to force that the output is always 1. This task is considered difficult if $C$ is extensive and is used as a basis of the RSA cryptosystem~\cite{rivest_method_1978}.
The second algorithm, \emph{Random 3-SAT}, is based on Sheridan et al.~\cite{sheridan_manufacturing_2016}. We generate a random conjunctive normal form (CNF) formula 
which consists of $32N$ clauses. The value of the factor $N$ is taken according to the previous results, which suggest $N \ge 6$ to have high chances for the clauses to be unsatisfiable. The formula is then negated, forcing the output always to be $1$.

Based on internal results, opaque predicates that used random 3-SAT yielded a worse size-to-performance ratio and were less robust against DSE. We evaluated our approach exclusively using the integer factorization algorithm for this reason.

\paragraph{Opaque Gadget Addresses Against Static Analysis}
\label{subsec:implementation:opaque:anti-rop-reversing}
We use opaque predicates to protect gadget addresses and immediate operands (Step \ref{obf-step:opaque} in Figure~\ref{fig:obfuscation-example}). An opaque predicate can generate a 1-bit output that is hard to compute statically.

For this reason, we concatenate each output bit of 32 distinct opaque predicate instances to generate a 32-bit constant (\emph{opaque constant}). We chose the arbitrary length of 32-bit for the constants since our work focused on a 32-bit architecture: naturally, this approach can be easily extended to different sizes.
In this way, static analysis attacks, whether automated or manual, need to reverse engineer the appropriate number of opaque predicates to compute the protected value.

\paragraph{DSE-resistant Opaque Predicates}
\label{subsec:implementation:opaque:anti-dse}
In the previous section, we discussed the generation of opaque predicates and their application to protect the gadget addresses against static analysis. However, this is not robust enough against DSE attacks (threat C). This section focuses on the steps we undertook to make the predicates DSE-resistant. We evaluated our approach with angr~\cite{shoshitaishvili_sok_2016}, but we believe it can be extended to other DSE engines. Concolic execution engines can execute code concretely; therefore, if the input to opaque predicates is statically known, the execution is deterministic. In this case, the engine can compute the opaque predicates' output very efficiently, resulting in the gadget addresses. The ROP chain would then be exposed and executed as if it were not obfuscated in the first place. However, if the input is unknown and concretized, the symbolic execution engine needs to evaluate the opaque predicates symbolically. This will force the computation of the mathematically complex problem on which the opaque predicate is based. Thus, we focused on finding appropriate input that cannot be easily concretized, imposing additional calculations on the symbolic execution engine.

\paragraph{Instruction Hiding}
\label{subsec:implementation:hiding}
Instruction hiding consists of the decomposition of an instruction into suboperations, of which, some are selected to be inserted out-of-order in other neighboring locations, protected by opaque predicates and the addition of dummy code. The reordering is applied exclusively to instructions whose order does not affect the outcome of the calculation.
We describe the process in more detail as follows.
\paragraph{Hidden Code Selection} In this step, the pass decides which part of the code is to be obfuscated by instruction hiding or by ROP transformation. It is essential to balance the two as an attacker can recover more instructions with execution tracing if ROP transformation is prevalent. On the other hand, the number of opaque predicates containing hidden instructions would be lower in the opposite case, leaving such instructions unprotected. We set a limit on the number of hidden instructions to be at most half of the total.
\paragraph{Embedding Code into Opaque Predicates}
The code is embedded into the opaque predicates through several insertion points.
The insertion points are designed to minimize register and flag conflicts with the inserted code. However, if there is a clash, we use temporary registers to preserve semantics, as explained in \S\ref{subsec:implementation:rop}.
\paragraph{Dummy Code Insertion} As a final step, we inject dummy code to the remaining insertion points to diversify the opaque predicates, avoiding trivial pattern-matching detection. Additionally, the dummy code is intertwined with code crucial to the computation of the opaque predicate. Besides useless computing operations, the code modifies the predicate's internal state and its variables, leading to confusion for the attackers.

\section{Evaluation}
\label{sec:evaluation}

\newenvironment{TakeAway}[1]{%
	\begin{tcolorbox}[arc=0pt,boxrule=0.5pt,coltitle=black,colback=black!5!white,colbacktitle=black!10!white,left=2pt,right=2pt,top=2pt,bottom=2pt,titlerule=0pt,enlarge top by=3pt, title=\textbf{Takeaway - {#1}}.]%
		\footnotesize}{%
	\end{tcolorbox}%
}

\def\ObfPlain{Baseline}
\def\ObfRoponly{ROPonly}
\def\ObfOpaqueBase{ROP+OP$_{\mathrm{Basic}}$}
\def\ObfOpaque{ROP+OP$_{\mathrm{DSE}}$}
\def\ObfHiding{ROP+OP$_{\mathrm{DSE}}$+Hiding}
\def\ObfBalanced{Balanced}

In this section, we evaluate \ropfuscator{} by addressing the following research questions.

\begin{itemize}[leftmargin=*,noitemsep]
	\item \textbf{RQ1: Completeness}. What maximum code coverage can our methodology achieve when obfuscating commodity software?
	\item \textbf{RQ2: Performance}. To what extent this obfuscation technique affects performance?
	\item \textbf{RQ3: Correctness}. Are the semantics of the program preserved?
	\item \textbf{RQ4: Robustness}. How is the robustness of the obfuscation mechanism concerning our threat model?
	\item \textbf{RQ5: Practicality}. Is our approach applicable to real-world use cases?
\end{itemize}

We measure the obfuscated instructions' coverage  \emph{(RQ1)} and performance overhead \emph{(RQ2)} in \S\ref{subsec:evaluation:coverage}, and \S\ref{subsec:evaluation:performance} respectively. During the test runs, we observe whether the program is executed correctly \emph{(RQ3)} and we evaluate our approach robustness \emph{(RQ4)} against the attacks defined in \S\ref{subsec:obfuscation:threatmodel} in \S\ref{subsec:evaluation:robustness}. Finally, we discuss practicality \emph{(RQ5)} by applying \ropfuscator{} onto an open-source media player. 

We evaluate several obfuscation configurations throughout this section. In particular, we select the following configurations that showcase varying degrees of obfuscation:

\begin{itemize}[noitemsep,topsep=1pt,parsep=1pt,partopsep=1pt]
	\item \emph{\ObfPlain}: non obfuscated binaries
	\item \emph{\ObfRoponly}: ROP transformation only  
	\item \emph{\ObfOpaqueBase}: ROP transformation, along with basic opaque predicates
	\item \emph{\ObfOpaque}: ROP transformation, along with DSE-resistant opaque predicates 
    \item \emph{\ObfHiding}: ROP transformation with opaque predicates and instruction hiding
\end{itemize}

\paragraph{Test Sets and Experiment Environment}

We evaluate coverage, execution speed (throughput), and code size with two test sets, the \emph{SPEC CPU2017} (SPECrate Integer) C/C++ tests and a subset of \emph{binutils}' applications. We evaluated the robustness with a simple program that validates a user-provided buffer.
Unless explicitly stated, the test cases are compiled with no optimizations (\Code{-O0} flag), the gadgets are extracted from the 32-bit libc version 2.27-3ubuntu1, and the programs are executed in a virtual machine running Ubuntu 18.04 x86-64.

\subsection{Completeness: Obfuscation Coverage}
\label{subsec:evaluation:coverage}

We first evaluate the ratio of obfuscated instructions when the sole ROP transformation pass is enabled. \ropfuscator{} is more robust when instructions are transformed since static disassemblers do not reconstruct the code executed by ROP gadgets.

\begin{table*}[t]
	\centering
	\footnotesize
	\caption{Ratio of instructions obfuscated in SPEC CPU 2017 (SPECrate Integer) C/C++ test cases.}
	\label{tab:coverage-speccpu}
	\resizebox{\textwidth}{!}{
		\def\n#1{\scalebox{.9}[1]{#1}}
		\begin{tabular}{ll|rrrrrrrrr|r}
			\toprule
			Option       & Status                         & \n{perlbench} & gcc     & mcf     & omnetpp & \n{xalancbmk} & x264    & \n{deepsjeng} & leela   & xz      & W.AVG   \\\midrule
			\texttt{-O0} & Obfuscated                     & 74.16\%       & 76.67\% & 64.79\% & 65.28\% & 66.03\%       & 64.75\% & 68.18\%       & 68.51\% & 66.23\% & 72.20\% \\
			             & Unobfuscated (No gadget / reg) & 9.83\%        & 8.59\%  & 11.68\% & 7.88\%  & 8.31\%        & 13.50\% & 11.38\%       & 6.59\%  & 11.86\% & 8.90\%  \\
			             & Unobfuscated (Other)           & 16.01\%       & 14.74\% & 23.53\% & 26.84\% & 25.66\%       & 21.75\% & 20.44\%       & 24.90\% & 21.91\% & 18.90\% \\\midrule
			\texttt{-O3} & Obfuscated                     & 40.41\%       & 42.41\% & 29.08\% & 43.20\% & 39.34\%       & 26.89\% & 33.14\%       & 42.29\% & 32.02\% & 40.33\% \\
			             & Unobfuscated (No gadget / reg) & 10.50\%       & 7.71\%  & 13.33\% & 7.41\%  & 9.53\%        & 11.22\% & 13.33\%       & 9.47\%  & 11.73\% & 8.74\%  \\
			             & Unobfuscated (Other)           & 49.09\%       & 49.88\% & 57.59\% & 49.38\% & 51.14\%       & 61.89\% & 53.54\%       & 48.24\% & 56.25\% & 50.93\% \\\bottomrule
		\end{tabular}
	}
\end{table*}

Table \ref{tab:coverage-speccpu} presents the ratio of obfuscated instructions in SPEC CPU test cases with optimization options \Code{-O0} and \Code{-O3}. When compiler optimizations are disabled (\Code{-O0}), about 60--80\% of the instructions are obfuscated into ROP chains. However, when all optimizations are enabled (\Code{-O3}), the average ratio decreases to around 40\%. This is because x86 is a CISC architecture, which makes it difficult to translate multiple specialized instructions into ROP microgadgets. Therefore, avoiding compiler optimizations may be more beneficial.

About 7--12\% of the instructions were not obfuscated due to the absence of accessible registers or gadgets, noting that the optimization levels did not impact this metric. 
It is possible to decrease this gap by saving and restoring registers to allocate free temporary registers or linking an ad-hoc library to access more gadgets.

\begin{table*}[t]
	\begin{tabular}{c}
		\begin{minipage}{0.41\textwidth}
			\caption{Ratio of instructions obfuscated in binutils for different libc versions.}
			\label{tab:coverage-binutils}
			\resizebox{\textwidth}{!}{
				\begin{tabular}{llrr}
					\toprule
					libc version & Status                         & readelf & c++filt \\\midrule
					2.27-3       & Obfuscated                     & 77.24\% & 74.99\% \\
					ubuntu1      & Unobfuscated (No gadget / reg) & 11.80\% & 11.70\% \\
					             & Unobfuscated (Other)           & 10.96\% & 13.31\% \\\midrule
					2.27-3       & Obfuscated                     & 36.02\% & 26.07\% \\
					ubuntu1.2    & Unobfuscated (No gadget / reg) & 53.02\% & 60.62\% \\
					             & Unobfuscated (Other)           & 10.96\% & 13.31\% \\\midrule
					2.31-0       & Obfuscated                     & 82.69\% & 80.93\% \\
					ubuntu9      & Unobfuscated (No gadget / reg) & 6.35\%  & 5.77\%  \\
					             & Unobfuscated (Other)           & 10.96\% & 13.31\% \\\bottomrule
				\end{tabular}
			}
		\end{minipage}
		\begin{minipage}{0.01\textwidth}
			~
		\end{minipage}
		\begin{minipage}{0.55\textwidth}
			\centering
			\caption{Runtime slowdown and code size of obfuscated programs for binutils for each obfuscation algorithm.}
			\label{tab:performance-binutils}
			\footnotesize
			\resizebox{\textwidth}{!}{
				\begin{tabular}{l l rr rr rr}
					\toprule
					       &
					       & \multicolumn{2}{c}{absolute value}
					       & \multicolumn{2}{c}{ratio (Baseline=1)}
					       & \multicolumn{2}{c}{ratio (Roponly=1)}                                                              \\
					\cmidrule(lr){3-4}\cmidrule(lr){5-6}\cmidrule(lr){7-8}
					metric & obfuscation                            & readelf & c++filt & readelf & c++filt & readelf & c++filt \\\midrule
					time   & \ObfPlain                              & 0.39s   & 0.30s   & 1.0     & 1.0     & 0.09    & 0.01    \\
					       & \ObfRoponly                            & 4.23s   & 30.6s   & 11.0    & 102     & 1.0     & 1.0     \\
					       & \ObfOpaqueBase                         & 41.1s   & 337s    & 107     & 1118    & 9.7     & 11.0    \\
					       & \ObfOpaque                             & 66.4s   & 761s    & 172     & 2527    & 15.7    & 24.8    \\
					       & \ObfHiding                             & 57.1s   & 611s    & 148     & 2030    & 13.5    & 19.9    \\\midrule
					size   & \ObfPlain                              & 1.1MB   & 1.1MB   & 1.0     & 1.0     & 0.10    & 0.07    \\
					       & \ObfRoponly                            & 10.5MB  & 15.7MB  & 9.6     & 14.1    & 1.0     & 1.0     \\
					       & \ObfOpaqueBase                         & 895MB   & 1407MB  & 828     & 1269    & 86.6    & 89.7    \\
					       & \ObfOpaque                             & 1530MB  & 2411MB  & 1417    & 2175    & 148     & 154     \\
					       & \ObfHiding                             & 1283MB  & 2063MB  & 1188    & 1861    & 124     & 132     \\\bottomrule
				\end{tabular}
			}
		\end{minipage}
	\end{tabular}
\end{table*}

Next, we observe the obfuscation coverage when linking our test cases against different library versions.
Table \ref{tab:coverage-binutils} shows the ratio of instructions obfuscated for two programs in \emph{binutils 2.32}, while Table \ref{tab:performance-binutils} focuses on the performance overhead.

There is a coverage decrease, around 20--40\%, between libc versions \emph{2.27-3ubuntu1} and \emph{2.27-3ubuntu1.2}. Furthermore, the instructions that are not obfuscated due to missing gadgets increase by 50--60\% on average. This is due to libc \emph{2.27-3ubuntu1.2} missing a gadget\footnote{\Code{xchg eax, edx; ret}} widely used by \ropfuscator{}. As previously described, we rely on exchange gadgets to convert single instructions into a combination of microgadgets~\cite{homescu_microgadgets_2012}. In this case, if a similar gadget is unavailable, it is impossible to use microgadgets relying on registers that must be exchanged first. Therefore, ROP transformation is very sensitive to ROP gadgets' availability, per the results obtained when linking a different libc version (\emph{2.31-0ubuntu9}). 
A straightforward solution to this problem would be to find such a gadget elsewhere, for instance, by linking a library that contains this gadget to the target application.

\begin{TakeAway}{RQ1: Completeness}
	On average, \ropfuscator{} transforms about 60--80\% of the instructions into ROP chains. The number depends on the compiler optimization option and shows better coverage without optimization. The number also depends mainly on the library version from which the ROP gadgets are extracted, and selecting an appropriate library version ensures high coverage.
\end{TakeAway}

\subsection{Performance and Correctness}
\label{subsec:evaluation:performance}

This subsection focuses on the run-time slowdown and code size overhead introduced by \ropfuscator{}.

\begin{figure*}[t]
    \resizebox{\textwidth}{!}{
	   \input{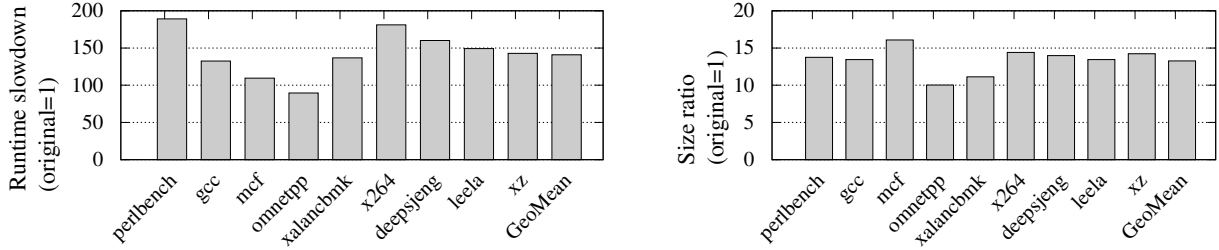}
    }
    \vspace*{-3\baselineskip}
    \caption{Runtime slowdown and code size of obfuscated programs for SPEC CPU 2017 (SPECrate Integer).}
	\label{fig:performance-speccpu}
\end{figure*}

In Figure \ref{fig:performance-speccpu}, ROP transformation causes a 140x (109-189x) mean execution time increase and a 13x (10-16x) binary size increase. Assessing opaque predicate obfuscation passes was unfeasible for SPEC CPU tests, so performance and code size overhead were evaluated on a binutils' applications subset.

Table \ref{tab:performance-binutils} reveals that opaque predicate obfuscation raises execution time by 10x without DSE resistance (\emph{\ObfOpaqueBase{}}) or 20x with DSE countermeasures (\emph{\ObfOpaque{}}). Code size increases by roughly 90x for basic opaque predicates and 150x for DSE-resistant ones. Instruction hiding (\emph{\ObfHiding{}}) reduces ROP gadgets, slightly improving performance with a 16x longer execution time and a 130x larger size.

Taking \emph{\ObfPlain{}} as a reference, the execution time is 10--200x longer with ROP transformation only, 200--4000x with ROP alongside DSE-resistant opaque predicates, and 150--3000x with all the obfuscation passes enabled. Simultaneously, the executable sizes are 10--16x bigger with ROP transformation only, 1500--2500x with ROP and DSE-resistant opaque predicates, and 1200--2000x with full obfuscation.

The overhead ratio is undoubtedly significant when obfuscating an entire program; however, heavy obfuscations are generally focused on specific portions of a program~\cite{borrello_hiding_particles_2021}. In the following \S\ref{subsec:evaluation:casestudy} we show how limiting the obfuscation does not impact the usability of real-world software. 

Finally, we observed no behavioral differences from the original programs during the experiments. We compared the obfuscated programs' output and checked whether it matched one of the unobfuscated versions. Though it is not formally proven to be correct, \ropfuscator{} could accurately preserve the program semantics in our experiments.

\begin{TakeAway}{RQ2-RQ3: Performance and Correctness}
	ROP transformation imposes 10--200x execution time and 10--15x code size overhead. DSE-resistant opaque predicates impose an additional 20x increase (total 200--4000x) in the execution time and a further 150x (total 1500--2500x) code size increment. While the results show significant overheads, controlling the performance/size overhead ratio is possible by carefully choosing critical locations that may be strongly obfuscated. A summary and robustness evaluation are shown in Table \ref{tab:tradeoff-summary}. Finally, \ropfuscator{} preserved the semantics of the tested programs throughout the obfuscation transformations.
\end{TakeAway}

\afterpage{
\begin{table*}[t]
	\centering
	\footnotesize
	\def\Breakable{\raisebox{-0.2ex}{\tikz\draw (0,0) circle (.8ex);}}
	\def\Robust{\raisebox{-0.2ex}{\tikz\draw[black,fill=black] (0,0) circle (.8ex);}}
	\def\RobustP{\raisebox{-0.2ex}{\begin{tikzpicture}\draw (0,0) circle (.8ex);\fill (0,.8ex) arc (90:360:.8ex) -- (0,0) -- cycle;\end{tikzpicture}}}
	\caption{Robustness and performance of each algorithm in \ropfuscator{} against attacks.}
	\label{tab:tradeoff-summary}
    \resizebox{\textwidth}{!}{
	\begin{tabular}{l cccc cc}
		\toprule
		\multirow{2}{*}[-2pt]{Obfuscation Algorithm}
		               & \multicolumn{4}{c}{Robustness against Attack Algorithm}
		               & \multicolumn{2}{c}{Performance}
		\\\cmidrule(lr){2-5}\cmidrule(lr){6-7}
		               & A) \ShortThreatA{}                                      & B) \ShortThreatB{} & C) \ShortThreatC{} & D) \ShortThreatD{} & Slowdown ratio & Size ratio \\\midrule
		\ObfPlain      & \Breakable                                              & \Breakable         & \Breakable         & \Breakable         & 1              & 1          \\
		\ObfRoponly    & \Robust                                                 & \Breakable         & \Breakable         & \Breakable         & 10--200        & 10-16      \\
		\ObfOpaqueBase & \Robust                                                 & \Robust            & \Breakable         & \Breakable         & 100--2000      & 900--1500  \\
		\ObfOpaque     & \Robust                                                 & \Robust            & \Robust            & \Breakable         & 200--4000      & 1500--2500 \\
		\ObfHiding     & \Robust                                                 & \Robust            & \Robust            & \RobustP           & 150--3000      & 1200--2000 \\
		\bottomrule
	\end{tabular}
    }
	\hfill \Breakable: Breakable, \Robust: Robust, \RobustP: Mostly Robust ~~
\end{table*}}

\subsection{Robustness}
\label{subsec:evaluation:robustness}
This section will refer to two programs whose source code is listed in Figure \ref{fig:source-crackme}. These examples represent simple input code validation routines that must be protected from the analysis.
We propose two versions of the source code, early and late exit, due to the different exploration behavior by symbolic execution engines when evaluating a return instruction. We evaluate robustness taking into account the threat categories highlighted in the threat model defined in \S\ref{subsec:obfuscation:threatmodel}: threats A \emph{(\ThreatA)}, B \emph{(\ThreatB)}, C \emph{(\ThreatC)} and D \emph{(\ThreatD)}. We used open-source decompilers to reverse-engineer the binaries: Ghidra, retdec, and r2dec. The results of this evaluation are summarized in Table \ref{tab:tradeoff-summary}.

\begin{figure}[htpb]
	\centering
	(a) early-exit function \hspace{3em} (b) late-exit function
\begin{lstlisting}[language=C, basicstyle=\scriptsize, numbers=left, numbersep=0pt]
 int check(const char *s) {      int check(const char *s) {                                        
                                    int i = 0;
   if (s[0] != 'H') return 0;       i += (s[0] == 'H');
   if (s[1] != 'e') return 0;       i += (s[1] == 'e');
   if (s[2] != 'l') return 0;       i += (s[2] == 'l');
   ...                              ...
   if (s[12] != '\0') return 0;     i += (s[12] == '\0');
   return 1;                        return i == 13;
 }                               }

         int main(int argc, char **argv) {
           if (check(argv[1])) puts("OK");
             return 0;
         }
\end{lstlisting}
	\vskip-0em
	\caption{Example of an input validating program.}
	\label{fig:source-crackme}
\end{figure}

\paragraph{Threat A: \ThreatA}

We evaluate robustness against this threat by decompiling the functions listed in Figure \ref{fig:source-crackme}. The code is compiled with and without obfuscations and processed by reverse engineering tools.

The results show that all decompilers can reconstruct the functions that have not been obfuscated successfully but fail in the presence of the ROP transformation pass. This is likely due to the decompilers' inability to correctly dissect the ROP chains' structure and recognize the function boundaries. The opaque predicates are decompiled successfully but require extensive human analysis to be removed from the decompiled code.

\paragraph{Threat B: \ThreatB}

We implemented a deobfuscator to dissect and extract the ROP chains to the original code, using a similar technique to deRop~\cite{lu_derop_2011}. We first identify the gadgets and later combine their underlying code into the original instructions. We do not process the ROP chains directly in memory; instead, our deobfuscation approach statically detects the ROP chain generator code by disassembling the binary, looking for stack manipulating instructions such as \Code {push}, \Code{pop} and \Code{ret}. Once the ROP chain is detected, we emulate the stack contents' execution, extracting the gadgets and concatenating their underlying code.
We deobfuscated binaries protected with the \ObfRoponly{} and \ObfOpaque{} configurations.

The recovered instructions from the \emph{\ObfRoponly{}} binary are slightly different from the original code, and the decompiled code has strong similarities and structure to the original C source code. On the other hand, our deobfuscator could not recover the code without fully emulating the opaque predicates' execution in the \emph{\ObfOpaque{}} binaries. Employing a CPU emulator diverges from this threat context as it would be more appropriate to Threat D, \ShortThreatD{}. Therefore, we believe that a targeted static ROP chain analysis can tackle the ROP transformation but cannot handle the opaque predicates.

Besides, we analyzed the immediate operands that appear in the ROP chains. In this case, we simulate a scenario where an attacker knows the application's constants. The example shown in Figure \ref{fig:source-crackme} is reflected by the comparisons in lines 3--7. Those statements are converted into a recognizable assembly pattern \footnote{e.g., \Code{push 0x48; push $G(\Code{pop ecx})$; \dots}} in the ROP chain, facilitating the extraction of the string constant. For this reason, we focused on extracting data from such patterns (i.e., periodic occurrence of immediate operands with various intervals). We successfully recovered the string constants in \ObfRoponly{} binaries but failed to do so from \ObfOpaque{} ones. This shows the effectiveness of protecting immediate operands with opaque predicates.

\paragraph{Threat C: \ThreatC}

For this scenario, we considered an automated attack for computing an input able to pass validation checks. We use the \emph{angr} symbolic execution engine to find such input that causes the application shown in Figure \ref{fig:source-crackme} to output `\Code{OK}.' We obfuscated the program with every obfuscation configuration.

\afterpage{
    \begin{table*}[t]
	\centering
	\footnotesize
	\caption{\ShortThreatC{} performance overhead for different obfuscation configurations and exploration strategies.}
	\label{tab:dse-time}
	\def\l##1{\raisebox{-1em}[0em][0em]{##1}}
     \resizebox{\textwidth}{!}{
	\begin{tabular}{ll rr rr rr rr}
		\toprule
		                                  &                                   &
		\multicolumn{8}{c}{\ShortThreatC{} exploration strategy in angr}
		\\\cmidrule(lr){3-10}
		\raisebox{0em}[0em][0em]{Program} &
		Obfuscation config                &
		\multicolumn{2}{c}{Symbolic/BFS}  &
		\multicolumn{2}{c}{Symbolic/DFS}  &
		\multicolumn{2}{c}{Tracing/BFS}   &
		\multicolumn{2}{c}{Tracing/DFS}
		\\\cmidrule(lr){3-4} \cmidrule(lr){5-6} \cmidrule(lr){7-8} \cmidrule(lr){9-10} &
		                                  & Time                              & Memory 
		                                  & Time                              & Memory 
		                                  & Time                              & Memory 
		                                  & Time                              & Memory 
		\\\midrule
		\l{Early-exit}                    &
		\ObfPlain                         & 5.4s                              & 170MB  
		                                  & 5.5s                              & 173MB  
		                                  & 4.5s                              & 170MB  
		                                  & 4.4s                              & 169MB  
		\\ &
		\ObfRoponly                       & 9.5s                              & 168MB  
		                                  & 8.0s                              & 164MB  
		                                  & 9.5s                              & 173MB  
		                                  & 7.4s                              & 176MB  
		\\ &
		\ObfOpaqueBase                    & 85.3s                             & 417MB  
		                                  & 57.4s                             & 365MB  
		                                  & 85.7s                             & 413MB  
		                                  & 56.1s                             & 368MB  
		\\ &
		\ObfOpaque                        & \multicolumn{2}{c}{Out of Memory}          
		                                  & \multicolumn{2}{c}{Out of Memory}          
		                                  & \multicolumn{2}{c}{Out of Memory}          
		                                  & \multicolumn{2}{c}{Out of Memory}          
		\\ &
		\ObfHiding                        & \multicolumn{2}{c}{Out of Memory}          
		                                  & \multicolumn{2}{c}{Out of Memory}          
		                                  & \multicolumn{2}{c}{Out of Memory}          
		                                  & \multicolumn{2}{c}{Out of Memory}          
		\\\midrule
		\l{Late-exit}                     &
		\ObfPlain                         & 4.5s                              & 130MB  
		                                  & 4.5s                              & 130MB  
		                                  & 3.7s                              & 130MB  
		                                  & 3.7s                              & 130MB  
		\\ &
		\ObfRoponly                       & 7.1s                              & 138MB  
		                                  & 7.3s                              & 134MB  
		                                  & 7.0s                              & 141MB  
		                                  & 7.0s                              & 141MB  
		\\ &
		\ObfOpaqueBase                    & 69.7s                             & 324MB  
		                                  & 68.2s                             & 326MB  
		                                  & 74.1s                             & 342MB  
		                                  & 74.0s                             & 345MB  
		\\ &
		\ObfOpaque                        & \multicolumn{2}{c}{Out of Memory}          
		                                  & \multicolumn{2}{c}{Out of Memory}          
		                                  & \multicolumn{2}{c}{Out of Memory}          
		                                  & \multicolumn{2}{c}{Out of Memory}          
		\\ &
		\ObfHiding                        & \multicolumn{2}{c}{Out of Memory}          
		                                  & \multicolumn{2}{c}{Out of Memory}          
		                                  & \multicolumn{2}{c}{Out of Memory}          
		                                  & \multicolumn{2}{c}{Out of Memory}          
		\\\bottomrule
	\end{tabular}
    }
	\begin{flushright}
		Out of Memory: force stopped after exceeding 8000MB ~~~~~~~~~~~~~~~
	\end{flushright}
\end{table*}
}

We employed different symbolic exploration strategies: depth-first vs. breadth-first search and symbolic vs. tracing. Additionally, we set a memory use threshold of 8GB and measured the time and memory needed to compute the input. The results are shown in Table \ref{tab:dse-time}.

As expected, \ShortThreatC{} is very effective in analyzing \ObfPlain{}, \ObfRoponly{} (less than 10 seconds) and \ObfOpaqueBase{} binaries (less than 2 minutes). On the other hand, it fails to find a valid input for \ObfOpaque{} binaries due to the exploration's significant memory requirements. Combined with DSE-resistant opaque predicates, ROP severely hinders input-finding attacks with symbolic execution techniques, even for a straightforward program like the one used in our tests.

\noindent\textit{User input in opaque predicates.} We observed that the sole application of a ROP transformation or opaque predicates independent from user input is not enough to protect against \ShortThreatC{}. As any branch based on user input values increases the time complexity of \ShortThreatC{}, we compute the gadget addresses using opaque constants intertwined with such input. We believe our approach should achieve similar results to the ones presented by Banescu et al.~\cite{banescu_code_2016}

Additionally, we propose a hypothesis that would challenge \ShortThreatC{}: frequently used general-purpose registers, such as \Code {eax} on the x86 architecture, might be used to handle user input.
We empirically verified our assumption by disassembling the compiled example code and noticed that each input byte is processed into the \Code {eax} register before being compared against an expected value.

This result indicates that our obfuscation measures introduced in \ref{subsec:implementation:opaque}, such as feeding user-supplied registers as input to opaque predicates, are effective against \ShortThreatC{}.

\paragraph{Threat D: \ThreatD}

Finally, we evaluate the robustness of \ropfuscator{} against \ThreatD{}. As previously noted, static analysis cannot efficiently analyze opaque predicates. For this reason, we emulate the ROP chain building code and subsequently implement a dynamic deobfuscator, adopting a similar approach to those proposed in previous works~\cite{graziano_ropmemu_2016,delia_static_2019}. The deobfuscator analyzes an execution trace generated by a CPU simulator for possible ROP chains. Furthermore, the trace is extracted from a code region provided by the user.

We applied the dynamic ROP deobfuscator to the early-exit function shown in Figure \ref{fig:source-crackme} (a) compiled with the configurations \ObfOpaque{} and \ObfHiding{}. We successfully extracted a semantically equivalent version of the code from the \ObfOpaque{} obfuscated binary. On the other hand, the code extracted from the binary obfuscated with \ObfHiding{} was only \textit{partially} matching the original one. This is due to instructions being hidden in the opaque predicates' body and, therefore, being ignored by the deobfuscator during analysis.

As a result, the sole use of opaque predicates is not robust against the dynamic tracing attack. Conversely, instruction hiding appears effective when combined with predicates, preventing code from being revealed in an execution trace.

\begin{TakeAway}{RQ4: Robustness}
	ROP transformation (\ObfRoponly) is robust against Threat A but not robust against Threat B, C, and D. Introducing opaque predicates (\ObfOpaque) fortifies the programs against Threats B and C. Finally, instruction hiding (\ObfHiding) makes the obfuscated binaries resistant against Threat D. The results are summarized in Table \ref{tab:tradeoff-summary}.
\end{TakeAway}

\subsection{Practicality: Case Study}
\label{subsec:evaluation:casestudy}

In \S\ref{subsec:evaluation:performance}, we evaluated the performance of \ropfuscator{} with hypothetical workloads, and the impact was significant. Such overhead is not compatible with any real-world application of our obfuscation technique. For this reason, we considered alternative options that retained a better performance overhead.
We assume that functions that produce or operate on sensitive data and, therefore, are subject to obfuscation interest take up a small portion of the total execution time. To balance robustness and performance, we considered obfuscating such functions selectively, and we later verified our assumptions in the rest of this section.

We apply \ropfuscator{} to an open-source media player, VLC. We explain how obfuscation mechanisms can be applied to protect critical assets in the program, balancing robustness and performance.

Digital rights management (DRM) is a mechanism to protect commercial media content against digital piracy~\cite{liu_digital_2003}. It is extensively used to protect video, audio, video games, and other media distributed on the Internet from unauthorized use.

The foundation of DRM is the encryption (or scrambling) mechanism to protect content. Attackers are interested in retrieving the encryption keys to decrypt the content to bypass the protection and illegally use or copy the material. Obfuscation plays an essential role in dissuading reverse engineering attempts that may disclose the algorithms and encryption keys employed by the DRM protections.

\paragraph{Protecting the DRM encryption routines}

We applied \ropfuscator{} to an open-source media player, VLC Media Player, using the DVD descrambling library \Code{libdvdcss}\footnote{\url{https://www.videolan.org/developers/libdvdcss.html}}. 
Upon VLC Media Player's request, \Code{libdvdcss} library derives the content decryption key (title key) from the protected DVD media and decrypts (descramble) the DVD content, which is later decoded and played by the media player on screen.

Though the final goal is to protect media content, we believe it is crucial to protect the title keys and the key derivation process. Therefore, we prioritized the obfuscation of the title key derivation more than the content decryption functions with the \emph{\ObfBalanced{}} configuration, ultimately aiming to achieve a better performance/size overhead ratio.

\paragraph{Evaluating the obfuscation impact} 

We obfuscated \Code{libdvdcss} with five configurations and then compared their performances:
\begin{itemize}[noitemsep,topsep=0pt,parsep=0pt,partopsep=0pt]
	\item \emph{\ObfPlain}, no alterations made to the code
	\item \emph{\ObfRoponly}, \emph{all} functions obfuscated with ROP transformation only
	\item \emph{\ObfOpaque}, \emph{all} functions obfuscated with ROP transformation and DSE-resistant opaque predicates
	\item \emph{\ObfHiding}, \emph{all} functions obfuscated with ROP transformation, DSE-resistant opaque predicates, and instruction hiding
	\item \emph{\ObfBalanced}, \emph{only title key derivation} functions obfuscated with \ObfHiding{} and the rest of the library with \ObfRoponly{}
\end{itemize}

\begin{table}[t]
	\centering
	\caption{Performance statistics of VLC Media Player using \Code{libdvdcss}.}
	\label{tab:libdvdcss-performance}
	\resizebox{0.5\textwidth}{!}{
	\begin{tabular}{lrrcr}
		\toprule
		Config       & Time  & CPU    & Played    & Size  \\[-0.2em]
		~            & [s]   & Usage  & Smoothly? & [MB]  \\\midrule
		\ObfPlain    & 30.2  & 12.4\% & Yes       & 0.034 \\
		\ObfRoponly  & 30.2  & 23.3\% & Yes       & 0.38  \\
		\ObfOpaque   & 110.2 & 97.0\% & No        & 48.5  \\
		\ObfHiding   & 120.7 & 95.3\% & No        & 41.3  \\
		\ObfBalanced & 30.2  & 23.2\% & Yes       & 18.4  \\
		\bottomrule
	\end{tabular}
	}
\end{table}
We tested each of these configurations by playing a commercial DVD title for 30 seconds\footnote{\Code{time vlc -I dummy dvd://\#1:3 --stop-time=745 vlc://quit}}, using VLC Media Player linked with \Code{libdvdcss}. The playback results are listed in Table \ref{tab:libdvdcss-performance}.

We measured the number of function calls and instructions using Valgrind~\cite{nethercote_valgrind_2007}. \Code{libdvdcss} accounts for about 5.5\% of the total instructions executed. Among \Code{libdvdcss}, the main DVD decryption function, \Code{dvdcss\_unscramble}, accounts for 99.88\% of the instructions (15091 out of 15970 function calls), while the key derivation functions only for 0.028\% of the instructions (63 function calls). Therefore, using a slower albeit more robust obfuscation for 0.0015\% of the overall instructions (0.028\% of \Code{libdvdcss}) does not significantly impact performance.

The average CPU usage is calculated by dividing the CPU utilization time by the total execution time\footnote{~$(T_{\mathrm{user}} + T_{\mathrm{system}})/T_{\mathrm{real}}$}. The results show that, without obfuscation, the program plays the DVD video smoothly, with the CPU usage averaging around 10\%. The application still plays the video smoothly when the sole ROP transformation is applied, albeit with an additional CPU 10\% usage on average. Applying opaque predicates upon the ROP transformation renders the media not playable in real-time, as the player frequently stops to buffer the movie contents and to peak the CPU usage to almost 100\%. Finally, using the \emph{\ObfBalanced{}} profile shows that the obfuscation overhead is almost on par with the sole ROP transformation while retaining the real-time decoding of the media.

These results seem to support our assumptions: retaining a reasonable performance/size ratio when obfuscating applications by targeting sensitive functions is possible. In this case, we could fortify the encryption routines while retaining the intended user experience.

\section{Discussion}
\label{sec:discussion}

In the previous section, we evaluated our work's performance, robustness, coverage, and correctness and demonstrated a real-world application of our technique.

We also demonstrated that \ropfuscator{} is robust against modern reverse engineering methodologies as defined in the threat model and, on average, can protect 70\% of the code (specifically, 60\%--80\% according to \S\ref{subsec:evaluation:coverage}). 

Our experiments highlighted the trade-off between the performance and robustness of our approach (Table~\ref{tab:tradeoff-summary}). However, tuning the obfuscation layers can be balanced per function (\S\ref{subsec:evaluation:casestudy}).

Although \ropfuscator{} is not specifically designed for data obfuscation, it can protect constant values by leveraging other techniques to work synergically with our approach. In the current implementation, we chose opaque constants to protect immediates in the binaries generated by \ropfuscator{}.  Moreover, users can embed and protect sensitive data such as constants used in white box cryptography~\cite{gu_white-box_2016} and recent developments in Mixed-Boolean Arithmetic (MBA) obfuscation~\cite{schloegel2022loki} suggest other promising ways to strengthen ROP obfuscation.

\paragraph{Portability of the obfuscated program}

Extracting gadgets from commonly used libraries, seemingly the natural path to follow, makes the obfuscated binary locked on the specific library chosen at compile time and, subsequently, not portable. For this reason, it is recommended to choose a library that ships with the program itself or create one from the ground up. In lieu of future work, we developed an accessory library called \textsc{librop}{\footnote{\url{https://github.com/ropfuscator/librop}}} providing user-specified gadgets and removing the limitations of using system libraries. 

Another solution to having a fully portable program is to link the library used to extract the gadgets statically.

\paragraph{On the use of microgadgets}

We adopted microgadgets for design simplicity and to speed up the prototyping cycle(\ref{subsec:implementation:rop}). While microgadgets are easier to extract, mapping a single instruction to one or more microgadgets skews the performance of the obfuscator. Spatial and time overhead depends on the number of instructions of a gadget and the number of gadgets needed to replace a specific instruction. The more gadgets are needed to replace a certain instruction, the higher the overhead. This characteristic aspect of ROP is an interesting research direction and should be pursued further as it could significantly impact the technique's overall performance. 

\paragraph{On the support of additional architectures}

ROPfuscator is designed for the 32-bit x86 instruction set as a prototype. Although supporting other architectures is achievable with further development, the core research concept remains unaffected by the present implementation.

\paragraph{Practicality}

ROPfuscator can be employed as a \textit{selective} obfuscation technique for obfuscating specific sensitive functions. However, its application to the entire code base is impractical due to the severe performance restrictions. Consequently, we advocate for exploring alternative directions while improving ROPfuscator key components' implementation, such as researching the impact of variable-length gadgets.

\paragraph{Resistance to other deobfuscation approaches}

First, we implemented a countermeasure to opaque predicate identification attacks~\cite{ming_loop_2015,tofighi-shirazi_defeating_2019}. We could not test the previous works for different reasons. We encountered technical issues in compiling the project by Ming et al.~\cite{ming_loop_2015}, that promptly offered us support. Unfortunately, the issues persisted.  Differently, we could not evaluate against the work by Tofighi-Shirazi et al.~\cite{tofighi-shirazi_defeating_2019} and Yadegari et al.~\cite{yadegari_generic_2015} due to the code being unavailable or not functioning at the time of writing.

We also examined VMHunt~\cite{xu_vmhunt_2018}, a virtualization deobfuscation method that claims compatibility with ROP chains. Regrettably, it was unable to identify any ROP gadgets in our experiments.

Furthermore, we tested Syntia, a program synthesis-based deobfuscation framework. It proved effective with simple functions but faced difficulties with more complex ones, irrespective of obfuscation, making it less suitable for our approach.

Finally, we considered a scenario where an attacker has complete knowledge of \ropfuscator{} and its mechanics. Although it is impossible to completely prevent an attacker from tracing ROP chain execution, instruction hiding offers some protection by disrupting the execution trace. This aligns with our objective of increasing analysis time cost.

\section{Related Work}
\label{sec:relatedwork}

This section briefly discusses the related studies on obfuscation and ROP. Moreover, we explain their relation to our approach.

\paragraph{ROP-based obfuscation}

ROP is applied in various ways to protect software. Recently, Borrello et al.~\cite{borrello_hiding_particles_2021} proposed an obfuscation approach based on ROP chains protected with a custom algorithm. 
This approach is based on static binary rewriting, i.e., a set of techniques to modify existing executable programs without the aid of a compiler and access to source code. As such, the developer must overcome challenges not present when hooking into a compiler, for instance, computing register liveness or finding a function's boundaries. Binary approaches are generally more error-prone and less resilient than techniques based on the compiler's visibility over the object files~\cite{sok:sp20}. However, we acknowledge that these approaches are more generic as they can be applied to closed-source software. 

Another common application is to evade detection from signature-based software such as anti-virus solutions. Frankenstein~\cite{mohan_frankenstein_2012} extracts ROP gadgets from benign binaries and combines them to generate ROP chains that execute malicious actions. However, this approach is not robust enough to counter an attack in a MATE scenario since it is not designed to withstand targeted ROP analyses. ROP needle~\cite{borrello_rop_2019} uses a similar technique to evade anti-virus detection by encrypting and decrypting the ROP chains on-the-fly using an externally supplied encryption key. ROP needle fits more a malicious scenario where malware authors intend to protect their work from analysts for a determined time frame (e.g., the duration of a malware campaign). However, there is no specified time limit for reverse engineering in commercial software protection, exposing the encryption key to an eventual disclosure to malicious analysts.

Another application is software tampering. Parallax~\cite{andriesse_parallax_2015} proposes a mechanism to embed ROP gadgets into sensitive code regions. Modifying these code regions leads to the ROP chain being corrupted, impeding its proper execution. Since they inherently change the program code, this can deceive debuggers when setting software breakpoints. Therefore, Parallax focuses on protecting software integrity and not its confidentiality.

In conclusion, several mechanisms protect software in MATE scenarios, sharing the objectives defined in our work~\cite{lu_ropsteg_2014,mu_ropob_2018}. RopSteg~\cite{lu_ropsteg_2014} proposes an instruction steganography algorithm. It injects ROP chains in code regions along with extra bytes. As an effect, this causes disassemblers to disassemble instructions erroneously. However, this considers only static analyses (Threat A), excluding a targeted ROP chain analysis executed through dynamic tracing (Threat D). ROPOB~\cite{mu_ropob_2018} exclusively obfuscates the control flow of a program, leaving non-control instructions unobfuscated, and its threat model does not consider targeted ROP analyses (Threat B and D).

\paragraph{ROP Chain Generation}

Q~\cite{schwartz_q_2011} proposes an approach to generate ROP chains. It defines semantic operations for branches, memory load/store, and arithmetic calculations. Later, it extracts and lifts the gadgets into an intermediate language that is finally compiled into a ROP chain. This technique is effective in generating ROP chains. We share this trait due to our ROP transformation pass, although its final goal is not obfuscation and thus has no discussion about preventing reverse engineering.

\paragraph{Opaque Predicates and Opaque Constants}

We use opaque predicates based on previous work. There is a multitude of algorithms proposed to generate them that span across various calculations, including arithmetic operations, non-determinism~\cite{collberg_manufacturing_1998}, one-way functions~\cite{zobernig_when_2019}, and computationally hard problems (e.g., pointer-aliasing~\cite{collberg_manufacturing_1998} or 3SAT~\cite{moser_limits_2007,sheridan_manufacturing_2016}). Furthermore, three main types of opaque predicates are documented in literature: invariant, contextual, and dynamic opaque predicates. \emph{Invariant} opaque predicates always evaluate to the same value, decided a priori by the obfuscator. \emph{Contextual} opaque predicates change their output based on preconditions chosen at design-time \cite{drape_intellectual_2010}. \emph{Dynamic} opaque predicates introduce the idea of correlated predicates that map a predicate's output to the input of a subsequent one \cite{palsberg_experience_2000}. 

These proposals are orthogonal to our approach, i.e., we can enhance \ropfuscator{} by integrating them as a component in our obfuscation. For this reason, we considered using opaque constants to compute gadget addresses~\cite{moser_limits_2007}.

\section{Conclusion}
\label{sec:conclusion}

We present \ropfuscator{}, a compiler-driven obfuscation pass based on ROP for any programming language supported by LLVM.
Although previous work already explores the effectiveness of ROP as an obfuscation technique, our approach deals with evolving reverse engineering attacks by introducing a unified threat model for ROP-based obfuscation techniques. We introduce a novel instruction hiding technique, later integrated with opaque predicates and constants, to provide a configurable yet robust framework.

The arms race between software obfuscation and reverse engineering seems to be endless. We introduce a unified threat model and a thorough evaluation along multiple dimensions to help us reason about the decisions involved in designing or choosing obfuscation techniques. This is an effort to provide researchers and practitioners with a better understanding of the strengths and limitations of obfuscation mechanisms. 

Finally, we release the source code of \ropfuscator{}\footnote{\url{https://github.com/ropfuscator/legacy}} to encourage future research in the systematic hardening of obfuscation schemes.

\section{Acknowledgements}

The authors want to express their sincere gratitude to John Ericson, Ilya Grishchenko, Francesco Mecca, and Fabio Pagani for their invaluable contributions, guidance, and support throughout the development of this study.

\bibliography{ropfuscator}

\end{document}